\documentclass[11pt,preprint]{aastex}

\shorttitle{Astronomy With Small Telescopes}
\shortauthors{Paczy\'nski}

\begin{document}

\title{Astronomy With Small Telescopes}

\author{Bohdan Paczy\'nski}
\affil{Princeton University Observatory, Princeton, NJ 08544}
\email{bp@astro.princeton.edu}

\begin{abstract} 

The All Sky Automated Survey (ASAS) is monitoring all sky to about 14 mag
with a cadence of about 1 day; it has 
discovered about ${\rm 10^5 }$ variable stars, most of them new.
The instrument used for the survey had aperture of 7 cm. 
A search for planetary transits has lead to the discovery of about
a dozen confirmed planets, so called 'hot Jupiters', providing the
information of planetary masses and radii. Most discoveries were
done with telescopes with aperture of 10 cm.

We propose a search for optical transients covering all sky with
a cadence of 10 - 30 minutes and the limit of 12 - 14 mag,
with an instant verification of all candidate events.
The search will be made with a large number of 10 cm instruments,
and the verification will be done with 30 cm instruments. 

We also propose a system to be located at the ${\rm L_1 }$ point 
of the Earth - Sun system to detect 'killer asteroids'. With a
limiting magnitude of about 18 mag it could detect 10 m boulders
several hours prior to their impact, provide warning against
Tunguska-like events, as well as to provide news about spectacular
but harmless more modest impacts. 

\end{abstract}  


\section{Introduction}  

The goal of this paper is to point out that there are many tasks
for which small and even very small telescopes are not only useful,
but even indispensable. Following are several examples. 

Gaustad et al. (2001) work on the ${\rm H_{\alpha} }$ emission was 
fundamental to account for Galactic foreground, therefore it was essential
for cosmology (Finkbeiner 2003); it was done with a 52 mm lens.
The same instrument was used to map a planetary nebula Abell 36,
which was found to be far more extended than thought previously
(McCullough et al. 2001). Optical flashes from cosmological distance
were detected by Akerlof et al. (1999, 
ROTSE - Robotic Optical Transient Search Experiment) and Jelinek et al. (2006):
they reached 8.9 mag and 10.1 mag, respectively, and used 100 mm 
optics. The flashes were due to gamma-ray bursts 990123 and 060117,
respectively. The recent optical flash at about 5 mag had unknown
origin (Shamir and Nemiroff 2006). It was registered with two
all sky cameras with 8 mm fish-eye lenses: one at Cerro Pachon - Chile,
another at La Palma - Spain. The flare was not detected at Cerro Paranal,
but the exposures were not coincidental, so there is no inconsistency.

Schaefer (1989) and Schaefer at al. (2000) complied a list of very
compelling cases where historical evidence implied unusual brightening
of otherwise ordinary bright stars. Unfortunately, none of these cases 
were rapidly followed up spectroscopically or photometrically.
There was no rapid followup in the recent case of Shamir and Nemiroff.
It is clear that without instant follow-up all those events remain
at the level of gossip.

However, for a number of years optical afterglows were successfully
followed up by many groups. The difference was due
to gamma-ray bursts acting as a trigger, making it possible to concentrate
on a relatively small area of a sky. If all sky was well covered
down to some magnitude, and the variability of all stars were known,
the recognition of transients would be much easier. It is one
of the goals of this study: to make the whole sky familiar at
gradually fainter magnitudes. This task is familiar: similar 
goals are advocated by LSST (Clever at al. 2004,
Large Synoptic Survey Telescope),
PanSTARRS (Kaiser 2004, Panoramic Survey Telescope \& Rapid Response System),
and SkyMapper (Schmidt at al. 2005, The SkyMapper Telescope). 
The major difference is the cadence: the three mega-projects
propose imaging all sky in a week. We suggest to image the 
sky once every minute or once every 15 minutes, depending how
deep we would go. Of course, all the proposed
mega-projects will reach vastly fainter magnitudes, 
and typically they will saturate below 15 mag. 
The huge difference in cadence and the 
magnitude make these projects complementary rather than competitive. 

In the following sections we shall present several examples of a
successful implementation of small instruments. ASAS
(All Sky Automated Survey) and NSVS (Northern Sky Variability Survey),
with the range of about 14 mag and cadence of about 1 day
provide an example of relatively low accuracy survey.
A search for planetary transits which concentrate an a very
small part of the sky, achieves a very high precision photometry.
These will be covered in sections 2 and 3. In addition we
propose two future project: the search for all untriggered
optical flashes, and a new approach of dealing with 'killer asteroids';
these are covered in sections 4 and 5. Finally, section 6 will conclude the 
paper with a discussion. 

\section{ASAS and NSVS}

The All Sky Automated Survey (ASAS) is an example of a modest system which
already exists, and it was it was to fill a gap in the sky variability.
It already produces results: catalogs of bright variable stars
(Pojma\'nski 1997, 1998, 2000, 2002, 2003, Pojma\'nski 
and Maciejewski 2004, 2005, Pojma\'nski, Pilecki and Szczygiel 2005).

ASAS is a long term project dedicated
to detection and monitoring variability of bright stars. 
It uses telescopes with the aperture of 7 cm, the focal length of 20 cm,
${\rm 2K \times 2K }$ CCD cameras with ${\rm 15 \mu }$ pixels
from Apogee. The standard V-band and I-band filers are used.
The I-band data are still being processed but all V-band data have
already been converted to catalogs of variable stars.
ASAS reaches 14-mag stars in 2 minute exposures over a field of view
${\rm (9^{\circ})^2 }$ degrees. 
More information about ASAS is provided on the WWW
\footnote{ http://www.astrouw.edu.pl/\~~gp/asas/asas.html \\ 
 http://archive.princeton.edu/\~~asas/ }
.
ASAS has already been used to address some problems in
the domain of contact binaries (Paczy\'nski et al. 2006).

In addition to discovering over 50,000 variable stars in the 
V-band, ASAS has also limited ability to react to new
phenomena: it discovered two comets, and a number of novae and dwarf
novae stars. However, the software has not been developed yet for a fully
automatic recognition of new phenomena.  Its main virtues are a
low cost and reliability - it requires Pojma\'nski to visit the
Las Campanas Observatory just once every two years.  The OGLE
(The Optical Gravitational Lensing Experiment) 
\footnote{ http://www.astrouw.edu.pl/\~~ogle/ \\
 http://bulge.astro.princeton.edu/\~~ogle/}
observer which happens to be on duty opens the dome for the OGLE
telescope, and this automatically opens ASAS.  About once every
week the OGLE observer changes data tape, or reboots the ASAS computer.

Pojma\'nski has recently (June 1, 2006) developed ASAS-N at Faulkes North
site at the Haleakala on Maui (Hawaii), courtesy of Wayne Rosing,
to cover all sky with the observations.  The optics has been upgraded 
to the aperture of 10 cm, with the same focal length. Just as the
ASAS at Las Campanas the system uses two cameras, one
for the V-band, another for the I-band.

The idea is to have ASAS as a system to operate indefinitely
as stars vary on all time scales. The practical issue is
to keep the operating costs very low, so the project can
continue for ever, with modest upgrades from time to time.
Hopefully, other astronomers, perhaps even amateurs, will
join ASAS-like projects, expanding it to all time zones,
and providing around the clock time coverage. 

The Northern Sky Variability Survey (NSVS,
Wozniak et al. 2004) provided a database of photometric
measurements covering time interval of 10 months of ROTSE-I (Robotic Optical
Variability Survey) data, and magnitude range 8 to 15.5.
While the original data were obtained with no filter, the
2MASS (Two Micron All Sky Survey) data can be used to obtain color information.
Combined with the current ASAS data the NSVS provides a 10-month
coverage of the variability of all sky. However, within a year ASAS will
expand its coverage of the Northern sky and will provide two
band photometry for the full sky. Still, the NSVS will remain
for ever as an archive: there is no way to go back in time. 

The usefulness of this activity can be quantified:
so far ASAS was used in various papers and notes listed
on ADS at least 134 times, the NSVS at least 52 times.

ASAS and NSVS make the sky well studied for variability down to
about 14 mag with a cadence of about one day. This may be a
good starting point to search for much more rapid transients,
with the better time coverage, and from many location. There
is no obvious limit to this activity.

\section{A search for planetary transits}

The search for planets transiting solar type stars produced spectacular
results.  The importance of transits is due to the fact that these are the
only planets for which accurate radii and masses can be obtained.
The first discovery was found using radial velocities to select
the candidate: HD 209458b (Mazeh et al. 2000, Charbonneau et al. 2000,
Henry et al. 2000). This was a bright star, suitable for detailed
photometry and spectroscopy. Two more cases of a transit selected
through radial velocity studies were found: HD 189733b (Bouchy et al. 
2005), and HD 149026b (Sato et al. 2005). These stars are very bright
too. 

So far 11 cases of planetary transits were discovered first
photometrically and confirmed spectroscopically later. 
Most of them were detected with 10 cm telescopes:
TrES-1 (Alonso et al. 2004), XO-lb (McCullough et al. 2006),
TrES-2 (O'Donovan et al. 2006), HAT-P-1b (Bakos et al. 2006),
WASP-1b and WASP-2b (Collier et al. 2006). 
I expect that the number of such discoveries will increase,
now that the astronomers learned how to make accurate photometry
with wide angle telescopes. The advantage of 10 cm
telescopes is that they cover large area in the sky, close
to ${\rm 100^2 }$ square degrees, and make a search broad.
Still, it will take many years to search the sky for planetary
transients. 

These are also 
5 objects first selected as candidates for planets by OGLE:
OGLE-TR-10b (Udalski et al. 2002a, Bouchy et al. 2005, Konacki et al.
2005), OGLE-TR-56b (Udalski et al. 2002b, Konacki et al. 2003),
OGLE-TR-111b (Udalski et al. 2002c, Pont et al. 2004),
OGLE-TR-113b (Udalski et al. 2002c, Bouchy et al. 2004, Konacki et al. 
2004), OGLE-TR-132b (Udalski et al. 2003, Bouchy et al. 2004).
All these are relatively faint as the stars were monitored with
a relatively large 1.3 meter telescope. Yet, for a year or
two OGLE was dominating the field of planetary transients
providing masses and radii for planets. 

%
Very recently, Sahu et al (2006) used the HST (Hubble Space Telescope)
to search for transiting planets at 19-26 mag; a survey known as
the Sagittarius Window Eclipsing Extrasolar Planet Search (SWEEPS).
Despite finding a few transiting candidates, spectroscopic follow-up and
confirmation of the SWEEPS candidates is currently out reach, since the
HST represents the extremely high end in both spatial and photometric
sensitivity.  Therefore, the SWEEPS planetary candidates must remain
candidates for some time come.\footnote{Of course, these are very
useful data about various types of variable stars, and in particular hundreds
of short period binary stars.}

It is clear that the best way to search for planetary transits is
to conduct it with many small telescopes, with an aperture
of 10 cm, or so, and large CCD cameras.  This is the best case I know
that the scientific advantage of small instruments is beyond any doubt.
Of course, there is a need to do spectroscopic follow-up with
bigger telescopes.

\section{A search for optical flashes}

The best known optical flashes are the gamma-ray burst afterglows (GRB).
A compilation of the results from GCN 
(The Gamma ray bursts Coordinates Network)
archive by Quimby and McMahon (2006) is available at 
WWW\footnote{ http://grad40.as.utexas.edu/\~~quimby/tu2006/}.
It indicates that a large fraction of optical afterglows decay
as ${\rm F \sim 1/t }$ initially, and as 
${\rm F \sim 1/t^2 }$ after the break.
Most afterglows are detectable only with large apertures,
though there were several very bright optical transients (OT):
OT 990123 (Akerlof et al. 1999, 9 mag), OT 060117 (Jelinek
et al. 2006, 10 mag), OT 061007 (Mundell et al. 2006, R=10.3 mag, 
Schady et al. 2006, V$<$11.1 mag). 

At this time the search for afterglows is frustrating, as most
of the time no afterglow is detectable, most time the instruments
are idle.  I think another mode of a search might be more satisfactory: 
to image all sky for whatever transients come around. There is
no fundamental rule that would restrict astronomy to just one
type of flashes. As far as I know this idea was seriously implemented
only by Robert Nemiroff and his associates with the concept of
CONCAM (ING's All-Sky Camera),
a camera with a fish-eye lens monitoring all sky in about
dozen observatories (Nemiroff \& Rafert 1999). Following several frustrating
years there was finally a success: the paper announcing the discovery
of a an optical flash of about 5th mag (Shamir and Nemiroff 2006).
The authors in their abstract were not suggesting an astronomical
discovery, even thought they recorded almost identical flush 
from Chile and from Canary Island.

Well, there was an obvious problem with the CONCAM: no automatic follow-up
that would provide a proof that the flash was real. The problem with
CONCAM, as well as all those stars of Schaefer (1989) and
Schaefer at al. (2000), was that there was no instant follow-up. 
It would be great to have 
spectroscopic verification, but a photometric verification should
be just as good, provided a larger instrument was available to
follow a promising candidate event down to 18th mag, or 
even down to 15th magnitude.
The obvious problem is to sort out which flashes were real and
which were artifacts. I am convinced that this problem can be
handled. After all these problems will have to be addressed with LSST,
PanSTARRS, and the SkyMapper, except they will be much tougher 
to handle with big instruments: there will be more candidate
flashes sampled very rarely, with a cadence of a week or so. 
The sky variability in a time domain is hardly explored. It makes sense to
explore the sky gradually, initially at the bright end, and
gradually to go deep.

The optical GRB afterglows are not the only optical sources
that may be discovered, as demonstrated by Shamir and Nemiroff
(2006) with their 5th magnitude flash. 
Systems similar to Nemiroff CONCAM
should be used to monitor the sky at all time scales. We do not
know what is the range of various optical flashes, GRB afterglows
and other phenomena. First, we should familiarize our system with
the sky at a given magnitude we can conveniently reach. As a by 
product we shell learn not only about the stars,
including all variables, but also asteroids, comets, etc., all kinds
of 'normal' transients.

Every instrument has a range of its applicability: magnitude range,
the cadence, sky coverage, etc. We have little information about
the best range of parameters to search, so we should search as
broadly as we can afford, in as many different parameters as 
we possibly can.  The new Apogee CCD cameras with 
${\rm 4K \times 4K}$ pixels 9 micron on a side cost about \$15,000.
Combining these cameras with telephoto lenses of different
foci's we can reach different magnitudes. For example, using
a lens with ${\rm f = 200 }$ mm we can reach down to 15 mag
in a few minute exposure. Taking faster optics and smaller
aperture we may reach brighter stars faster. For example, 
we can image all stars brighter than 10th mag every minute,
or so, providing 'continuous record of the sky' (Nemiroff
\& Rafert 1999).  This would make it possible to search the sky for optical
flashes with verification.  Gradually, improved optics will
allow, among other topics, to search for optical
afterglows without GRB triggers. 

\section{Killer asteroids} 


The search for so called `killer asteroids', with diameters in excess of
300 meters and Earth crossing orbits, is one of the most active areas of 
solar system research, with over 3500 objects discovered up to 2005-Dec
\footnote{ http://neo.jpl.nasa.gov/neo/number.html }.
The rate of collisions of 1 km asteroids with the Earth is uncertain.
It is estimated to be once every ${\rm 10^5 }$ years by 
Rabinowitz et al. (2000),
once every half a million years by Ivezi\'c et al. (2001), and once every
3 million years by Brown et al. (2002).  
Currently, there are approximately ten NEO (Near Earth Object) search programs
\footnote{http://neo.jpl.nasa.gov/programs/ }.
These discoveries provide information about possible or probable impacts
in a distant future, but none of them predicted any actual impact.

Collisions with ${\rm D = 1 }$ km asteroids are likely to be globally
catastrophic, but they are very rare according to all estimates.
Much more common are Tunguska-class events (Chyba et al. 1993), 
which occur every 1,000 years,
according to Brown et al. (2002), and more frequently according to
Rabinowitz et al. (2000) and Ivezi\'c et al. (2001).  While these are of no
global concern they would be locally catastrophic, equivalent to an explosion
of a major thermonuclear weapon, ${\rm \sim 10 }$ megatons of TNT, i.e. 500 
TIMES More powerful than the Hiroshima bomb.  Even far less energetic events 
could be locally catastrophic. More information about meteor phenomena 
can found in Ceplecha et al. (1998). 

As far as I 
know there is no ongoing search for such small, ${\rm \sim 10 - 50 }$ meter 
size objects, which could provide advance warnings of their approach and 
impact.  While even a Tunguska-class asteroid is not very likely to strike 
within a decade, it would be a major embarrassment for the astronomical 
community not to provide a timely prediction of a strike.

This is an outline of a project to
detect cosmic rocks several hours, perhaps several days prior to the impact,
and to provide an advanced warning to the local population 
(cf. Paczy\'nski 1997, 2000, 2001).  Smaller objects
might also be detected, and the prediction of the time and location of
harmless but spectacular fireballs could provide astronomical
entertainment to the general public (cf. Foschini 1998, Tagliaferri 1998).

Brown et al. (2002) estimated the flux of small near-Earth objects
(NEOs) colliding with the Earth based on the record of events
detected by the United States Department of Defense and Department of Energy.
The events were recorded with space-based systems with infrared detectors
in geostationary orbits (Defense Support Program - DSP).
The energetics of infrared flashes was converted to the kinetic energy of
the bolides and into their likely size.
In total, ${\rm \sim 300 }$ spectacles were recorded from February 1994 to 
September 2002, but none was predicted.

While Brown et al. (2002) and Ivezi\'c et al. (2001) give discordant estimates
for collision rate with ${\rm D \approx 1 }$ km asteroids, their estimates
for small, ${\rm d \approx 10 } $ meter rocks are similar.  The cumulative 
rate may be approximated as
$$
{\rm N \approx 0.1 \times \left( { D \over 10 ~ m } \right) ^{-2.5} ~ yr^{-1} ,
}
\eqno(1)
$$
i.e. the rocks 10 meter in size strike the Earth every 10 years, on average.
It takes ${\rm \sim 24 }$ hours to move a distance of 4LD (Lunar Distance)
at the velocity of ${\rm 20 ~ km ~ s^{-1} }$. 
This gives us little time to provide a warning.

The apparent magnitude of an object with diameter D located in the 
anti-solar direction at a distance d is estimated to be
$$
{\rm V \approx 18 -5 ~ \log
\left[  \left( { D \over 1 ~ km } \right)
 \left( { 1 ~ AU \over d } \right) \right] \approx
15 - 5 ~ \log
\left[ \left( { D \over 10 ~ m } \right)
\left( { 1 ~ LD \over d } \right) \right] = }
$$
$$
\approx 
{\rm 18 - 5 ~ \log
\left[ \left( { D \over 10 ~ m } \right)
\left( { 0.01 AU \over d } \right) \right] ,}
\eqno(2)
$$
where AU is an astronomical unit, which is approximately equal
to the inner Lagrangian point in the Earth - Sun system.
It is also the location of a space probe SOHO - the Solar and Heliospheric
Observatory (Fleck 2004).  The apparent magnitude can by estimated noticing
that full Moon is $ - 12.3 $ mag, and a rock with the 10 meter 
diameter at the Moon distance will be 15 mag, assuming Moon's albedo.

The best view for detection of a space rock is at a full Moon phase.
At a quarter Moon the brightness is 10 times smaller, and at new Moon
phase the rock is not visible at all.  Yet, Earth is under bombardment
all the time, from solar as well as anti-solar direction. There
is only one way to provide all time protection: go to space, preferably
to the L1 point, where all space rocks are well illuminated,
assuming we look at them from L1, and we monitor approximately
${\rm \pi }$ steradians. 
Adopting L1 point to look for the boulders aiming at Earth, we can
can detect a rock with a 10 meter diameter as a star of 
18 mag, with a modest change depending on its phase angle. A rock
one hundred meters across would be seen as a 13 mag star
from the inner Lagrangian point. 

Obviously, there are problems. The number of speeding rocks will be
huge, most of them will miss the Earth by a large margin. Some can
be rejected right away, but many will have to be followed to
make sure they miss the Earth. As the positions of all objects
will very all the time, it will be necessary to image them many 
times a day in order
to be able to keep track of them. The estimate is well beyond the
task of this paper. 

A very obvious question: will it be better to use a relatively
large telescope, say 50 cm diameter, to monitor 18th mag objects,
or will it be better to have a number of smaller instruments 
to do the task? In any case the sky
has to be imaged many times every day. Assuming
that monitoring will cover ${\rm \pi }$ steradians of the
sky centered on Earth as seen from the ${\rm L_1 }$ point, 
this will be a serious number crunching project, and even
more serious programming task. However, if we are serious 
about 'killer asteroids' this is the best way to handle the
danger. Notice, this is the most likely danger that may be
caused by astronomy, not stellar explosions or gamma-ray bursts. 

\section{Discussion}

The first three chapters: the introduction, the description
of ASAS and NSVS, and the description of planetary transit,
are the practical applications of small telescopes, which
generate interesting results. These will continue, and the
future of these small instruments is without any doubt.

The next two chapters are the expectations. The search
for optical transients is likely to be developed by Rosing,
Pojma\'nski and Paczy\'nski (2006) using small instruments.
The future results by LSST, PanSTARRS, and the
SkyMapper will be complementary, with their magnitude
range and a cadence of approximately one week.  

The proposed approach to 'killer asteroids' will require
much studies to evaluate the practical aspects of the proposed scheme.
I am optimistic. The search for on coming boulders
may provide considerable excitement, in particular those
which result in very bright optical flashes, like those
of Foschini (1998) and Tagliaferri (1998); except these flashes 
will be predicted in advance. 

\acknowledgments

It is a great pleasure to thank M. Collinge, A. Socrates and D. Szczygiel for
their help in various stages of this project. 
The author would also like to thank A. Filippenko for many useful comments.
NSF grant AST-0607070 and NASA grant NNG06GE27G are also acknowledged.

\clearpage 

\end{document}